# Emergence of Power-Law and Other Wealth Distributions in Crowd of Heterogeneous Agents

Jake J. Xia[1]

## Abstract

This study investigates the emergence of power-law and other concentrated distributions through a feedback loop model in crowd interactions. Agents act by their response functions to observations and external forces, while observations change by the aggregated actions of all agents, weighted by their respective influence, i.e. power or wealth. Agents' wealth dynamically adjust based on the alignment between an agent's actions and observation outcomes: agents gain wealth when their actions align with observed trends and lose wealth otherwise. A reward function, that describes the change of agents' wealth at each time step, manifests the differences of response functions of agents to observations. When all agents' responses are set to zero and feedback loop is broken, agents' wealth follow a normal or lognormal distribution. Otherwise, this response-reward iterative feedback mechanism results in concentrated wealth distributions, characterized by a small number of dominant agents and the marginalization of the majority. Contrasted to past studies, such concentration is not limited only to asymptotic behavior at the upper tail for large variables, nor does it require the reward function to be linear to agents' previous wealth as formulated in random growth model and network preferential attachment. Probability density functions for various distributions are more visually distinguishable for small values at the lower tail. In application of this model, key differences in income and wealth distributions in the US vs Japan are attributed to different response functions of agents in the two countries. The model's applicability extends beyond social systems to other many-body systems with analogous feedback mechanisms, where power-law distributions represent a rare subset of general concentrated outcomes.

---

[1] Harvard Management Company, Research Affiliate at Massachusetts Institute of Technology. Email: jjxia@mit.edu. This paper is a revised version of a 2010 working paper.

## I. Introduction

Background

Concentrated distributions are prevalent in many social and natural phenomena. The phenomenon of “rich-get-richer” or “winner-takes-all” is widely observed. For instance, wealth distribution in a country often demonstrates such concentration. In the field of venture capital investing, top-tier funds benefit from their previous successes, attracting more high-quality startups. These funds, in turn, leverage their networks and resources to better support startups, increasing their chances of success [Mallaby 2022].

This phenomenon, known in economics as the Matthew Effect [Merton 1968], describes how cumulative advantages allow agents’ power to grow proportionally with their initial level of power and networks, creating a feedback loop. In social media, a similar dynamic occurs: posts that gain more views and likes attract even more viewers, leading to viral dissemination.

Power-law distributions are often employed to fit data and determine the linear slope on log-log plots of expected value versus rank. While many studies focus on the slope of such charts, the underlying mechanisms causing power-law distributions are less well understood.

Power-law distributions contrast sharply with normal distributions, whose probability density functions form a bell shape. Normal distribution is often observed in nature, e.g. height distribution of adults. These types of data are usually observed in events with independent random variables without dynamic interaction. When there is no feedback loop, the variable changes randomly at each time step. After many time steps, the variable follows a normal distribution, as suggested by Central Limit Theorem, and probabilities peak around the mean of the random variables. In contrast, power-law distributions exhibit a peak at extreme values, often near the minimum boundary of the variables.

Power-law distributions represent a special case where a system reaches a stationary state and becomes self-similar at all scales, i.e. scale-free. Power-law distributions are inherently scale-invariant, a property unique among statistical distributions. When values in a power-law distribution are scaled, the resulting distribution retains proportional similarity to the original. This scale-invariance underscores the elegance and rarity of power-law distributions, as only distributions with power-law tails exhibit this property. In general, systems may continue to evolve beyond this steady state, often leading to further concentration and instability. This is why perfect scale-free networks or power-law distributions are rare [Broido 2019].

Literature Review

The emergence of power-law distributions has been studied across disciplines, providing insights into the mechanisms driving their formation. These distributions are often characterized by their "rich-get-richer" dynamics, where entities with more resources or connections are disproportionately likely to gain additional resources.

This phenomenon, also known as preferential attachment, has been widely observed in random networks [Barabasi 1999, Amaral 2000] and citation systems [Price 1976], where highly connected nodes attract more connections. Preferential attachment and network growth are the two critical causes to power-law distribution for the number of links between network nodes. After many time steps and the number of links becomes sufficiently large, the network links settle into a stationary scale-free state. The probability of attachment is assumed proportional to a node's links established up to that point of time divided by total number of links, which grows linearly over time. Power-law only occurs when effective boundary conditions are added to the linear reward function.

There are other mechanisms such as self-organized criticality, introduced by Bak [Bak 1987], describe systems evolving to critical states without external tuning. This framework has been pivotal in explaining power-law distributions in natural systems, including forest fires and sand pile avalanches. Jensen [Jensen 1998] expanded on these ideas by offering a detailed overview of self-organized criticality's role across various phenomena, highlighting its application in explaining cascading failures and large-scale systemic changes.

Fractal geometry and scale-invariance, as discussed by Mandelbrot [Mandelbrot 1982], also underpin the emergence of power-laws. These systems exhibit self-similarity across scales, resulting in structures like coastlines, riverbanks, and geophysical phenomena following power-law distributions. Turcotte [Turcotte 1999] further examined these patterns, linking hierarchical structures and growth constraints to the manifestation of power-laws in both natural and social systems.

The economics of superstars, as described by Rosen [Rosen 1981], illustrates the skewed reward dynamics in markets where minor differences in talent lead to amplified disparities in outcomes. This phenomenon, driven by network effects of global visibility and reputation, has been observed in fields like entertainment, sports, and technology, where a few individuals dominate the market share or audience base. In markets, advantages like brand recognition, network effects, and economies of scale lead to the disproportionate success of a few entities.

Optimization models [West 1997] provide another framework for understanding power-law distributions. Systems optimized for efficiency or resource allocation, such as transportation networks and biological systems, often exhibit scale-invariant solutions. Examples include energy-efficient blood vessel networks and optimized urban layouts, where hierarchical resource allocation results in fractal and power-law patterns. Zipf's law [Zipf 1949] in language emerges from optimization of communication, balancing speaker effort and listener understanding.

Extreme value theory offers a lens for understanding power-laws in rare, high-impact events. This theory has been applied to phenomena such as earthquakes and financial crashes [Gumbel 1958, Campbell 1997, Sornette 2003], where the tail behavior of distributions reveals the disproportionate likelihood of extreme outcomes. Mantegna's [Mantegna 1995] work on random walks and diffusion models further complements this understanding by linking constrained movements to skewed distributions observed in stock prices and animal foraging behaviors.

Evolutionary models and reinforcement dynamics, as explored by Solé [Solé 1996] and the Yule process [Yule 1925], provide insights into species diversity and population sizes. These models emphasize survival-driven dynamics and positive feedback loops, which result in power-law distributions. Simon's aggregation models further explain power-laws through merging and fragmentation processes, offering insights into city sizes [Simon 1955], firm growth, and linguistic patterns.

The interplay of random growth models [Gabaix 2009] with boundary conditions [Mantegna 1995] has been extensively studied to understand the conditions under which power-laws emerge. Random multiplicative processes with proportional growth lead to lognormal distributions. Power-law distributions in firm sizes appear as an asymptotic behavior in the upper tail of large values [Simon 1955, Sutton 1997, Mitzenmacher 2004]. The nuances of transitioning from lognormal to power-law distributions under specific constraints form the basis for understanding the random growth model.

Reinforced random walks [Davis 1990] use past movements to influence future behavior, such as website traffic or urban mobility. Unlike stationary systems, these often exhibit time-varying probability distributions, contingent on initial conditions and agent responsiveness. Highly reactive agents can synchronize collective behavior, as observed in crowd dynamics.

The multifaceted nature of power-law distributions in diverse fields encompasses mechanisms from preferential attachment to optimization, self-organization, and evolutionary dynamics. The fundamental causes of power-law distributions are closely tied to growth dynamics, feedback loops, and constraints, all of which shape complex systems. While exact microscopic interactions are often too complex to model directly in many-body systems, macroscopic models offer insights into system-level behavior. Such approach is commonly used in physics [Xia 1994].

The common thread in all the above studies is a positive feedback loop that derives power-laws in economic systems through network effects and economies of scale. This insight of the feedback mechanisms in complex systems links applications in finance, biology, physics, and social sciences.

With the rapid technological advances, the world is increasingly linked. More feedback loops are established with stronger responses. Such impact on society and economy is attracting more attention from top academic researchers [Acemoglu 2023].

## Focus of This Paper

Building on previous research [Xia 2016], this paper introduces a general framework to analyze the emergence of power-law distributions in crowd interactions. The feedback loop connects agents' actions to their observations. For the sake of generalization, this study uses the terms "crowd" and "system,", "agents" and "elements," as well as "power" and "wealth" interchangeably, to capture broad social and natural applications.

In earlier studies of crowd synchronization [Xia 2016], focus was on agents' heterogeneous response functions. Agents act differently were simplified to have homogeneous influence or "power" across the system. This study extends the model by incorporating heterogeneous wealth in agents with varying response functions. A reward function is introduced to describe how agents' wealth change over time. The objective is to derive a mathematical framework on how the reward of wealth is linked with agents' response functions. Emergence of power-law distributions of agents' wealth can be better understood. This framework conceptualizes a multi-agent crowd as a social analog to many-body systems. The feedback loop between agents' actions and observations often gives rise to trend-following behaviors, which lead to crowd synchronization. This synchronization, in turn, results in the emergence of "super-agents" who wield significantly higher power than other agents.

## II. Power-Law Distribution Compared with Other Distributions

This section explores the distinctions and relationships between power-law distributions and other statistical distributions, incorporating mathematical representations and illustrative examples.

### Representations of Power-law Distribution

A significant source of confusion in discussing power-law distributions stems from the variety of representations employed, including probability density functions (PDF), cumulative distribution functions (CDF), and expected value vs rank. Zipf's Law, for instance, describes the expected value of ranked variables as a function of rank, while Pareto's Law describes cumulative probability as a function of random variables. The most generic representation is the PDF, where the probability density *f(x)* is presented as a function of random variable *x*.

Normal distribution's PDF (probability density function) is a Gaussian function

$$f(x) \sim e^{\frac{-(x-\mu)^2}{2\sigma^2}}$$

PDF of power-law distribution is $P[X=x] = f(x) \sim 1/x^{a}$, where $ln(f)$ and $ln(x)$ have a linear relationship, $-a$ is the slope of a linear curve in log-log plot. PDF is a common way to discuss statistical behaviors, such as normal or lognormal distributions. If plotted, $y$-axis is the probability density function *f(x)* and *x*-axis is the random variable *x*.

CDF (cumulative distribution function) for power-law distribution, $P[X>x] = g(x) \sim 1/x^{k}$ , which is Pareto's Law. This is also called counter-cumulative distribution function contrasting with CDF of $P[X<=x]$. Cumulative probability can be calculated by integrating PDF $f(x)$ over the range of $x$. $f(x)$ can be calculated by taking the derivative of CDF. Hence, $a=1+k$.

Expected Value $E[x]$ vs rank $r$: $E[x]\sim 1/r^b$, which is Zipf's Law. If we plot, y-axis is the expected value of random variable $x$ and x-axis is the rank $r$ of $x$ from the largest to the smallest values.

The relationship linking the three representations in Zipf's law, Pareto's Law and PDF is, $k=1/b$, $a=1+1/b$ [Adamic 2000]. In our later discussions, we can freely switch between these representations.

Scale free (or scale-invariance, or self-similarity at different scales) (80-20 rule across whole range) is a unique feature of power-law. The mathematical proof is easy to find in textbooks or online. That is, "scale-invariance" can be thought of interchangeably with "power-law." The mathematical property of scale invariance is only achievable with a power-law distribution. This is because when you scale up (multiply) a value in a power-law distribution, the resulting probability distribution remains proportionally similar to the original.

The scale-invariance is an elegant property, but it is rare [Brodio 2019]. Distributions with "power-law tails," i.e., tails that match the Pareto distribution up to a multiplicative constant, are often cited as the only scale-invariant distributions. In later discussion, I will show that power-law can emerge for the whole distribution under certain circumstance.

**Comparison to Normal and Lognormal Distributions**

Normal distributions, often observed in nature, exhibit bell-shaped PDFs with probabilities peaking at mid-values. In contrast, power-law distributions belong to a broader class of concentrated distributions, characterized by probabilities that drop sharply from their peak as variable values increase, i.e. its first order derivatives being negative. Because the PDF is floored at zero, this leads to a convex (i.e. its second order derivative being positive) curve with probability density peaking at minimal values. The boundary condition of a minimum value is often necessary for power-law. Lognormal distributions serve as an intermediary, resembling power-law at large values while maintaining bounded minimal value.

The study of low-value tails in power-law distributions is particularly critical, as these regions represent the majority of agents within the system. For example, in income and wealth distributions, a normal distribution suggests a majority of middle class in the population, typically known as the American football shape distribution. A power-law distribution indicates a concentration of wealth among a small elite, with the majority of individuals situated at minimal wealth levels.

Both normal distribution and power-law distribution have very large values of the random variables, e.g. extremely rich people as small percentage of the population. The difference is where majority of people sit. Concentrated distributions usually have probability peaking at minimum values. Traditionally, data collection and statistical studies have focused on ultra-rich of top 0.1% to 10%. Studying lower tail of small values (poor agents) is more important than the upper tail of large values (rich agents) to differentiate power-law from normal or lognormal distributions.

Figure 1a. PDF of Distributions

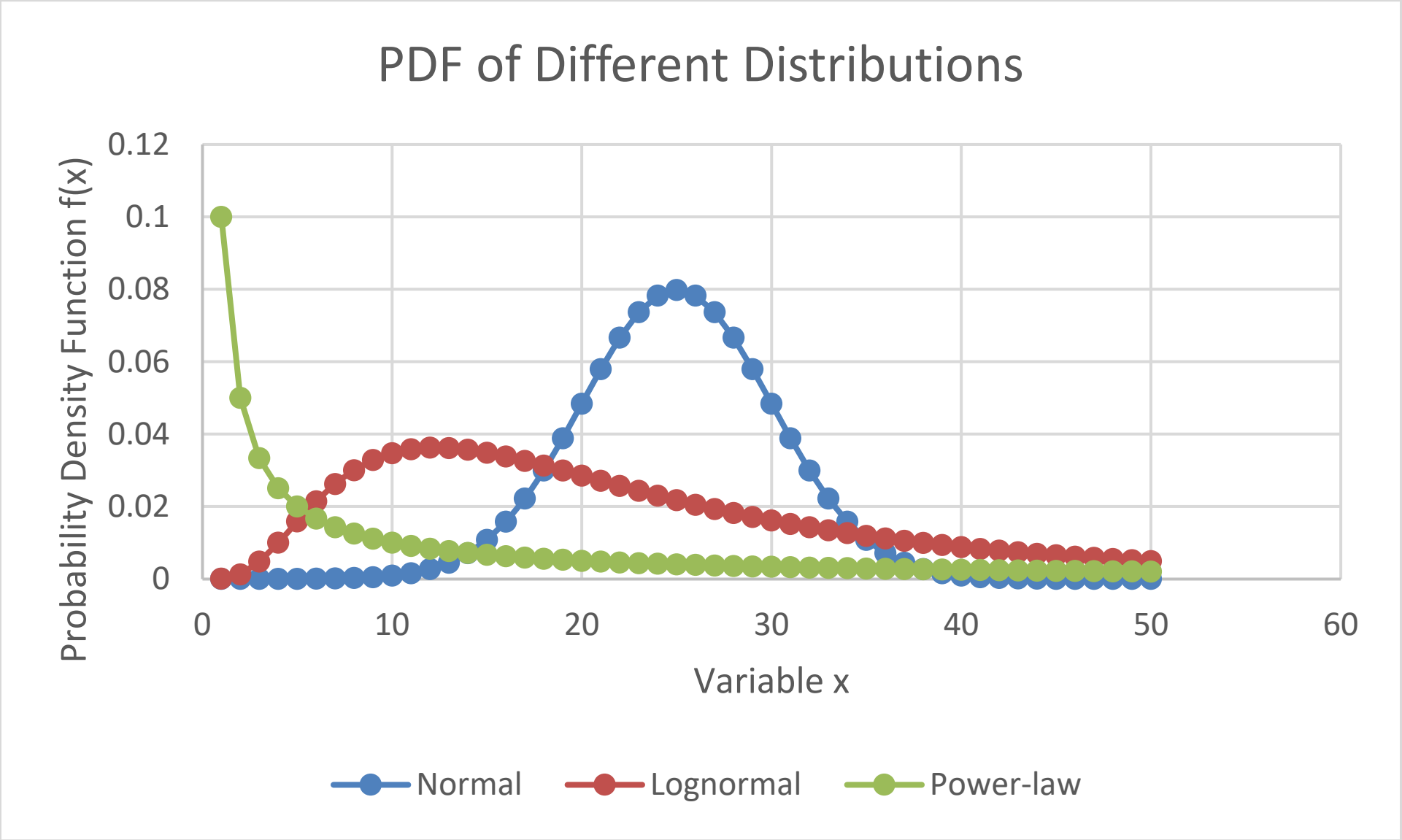

Figure 1a shows that it is easy in PDF to distinguish power-law from normal and lognormal distributions by looking at lower tails.

**Expected Value as a Function of Rank**

Power-law distributions are often shown in rank space with agents' rank as the x-axis and expected value of the variables as y-axis. For wealth distribution, we can plot with y-axis as wealth and x-axis as the population percentile from the richest to the poorest. For even distribution, the curve is a flat horizontal line. In order to compare expected values of different ranks, the rank buckets need to be evenly spaced. For example, if data is collected for total wealth owned by top 1% vs bottom 50%, one needs to covert the data to wealth per household or per person.

To show expected value $E[X=x]$ vs rank for normal distribution involves bucketing all $x$ values by equal spacing of probability (e.g. percentiles) and then rank them. For example, rank 1 is 90$^{th}$ percentile, rank 2 is the 80$^{th}$ percentile, and so on. Equal spacing means that we bucket the same number of units in each rank. Expected value is the median or probability-weighted value in each bucket.

Figure 1b. Distributions Presented in Percentile Ranks

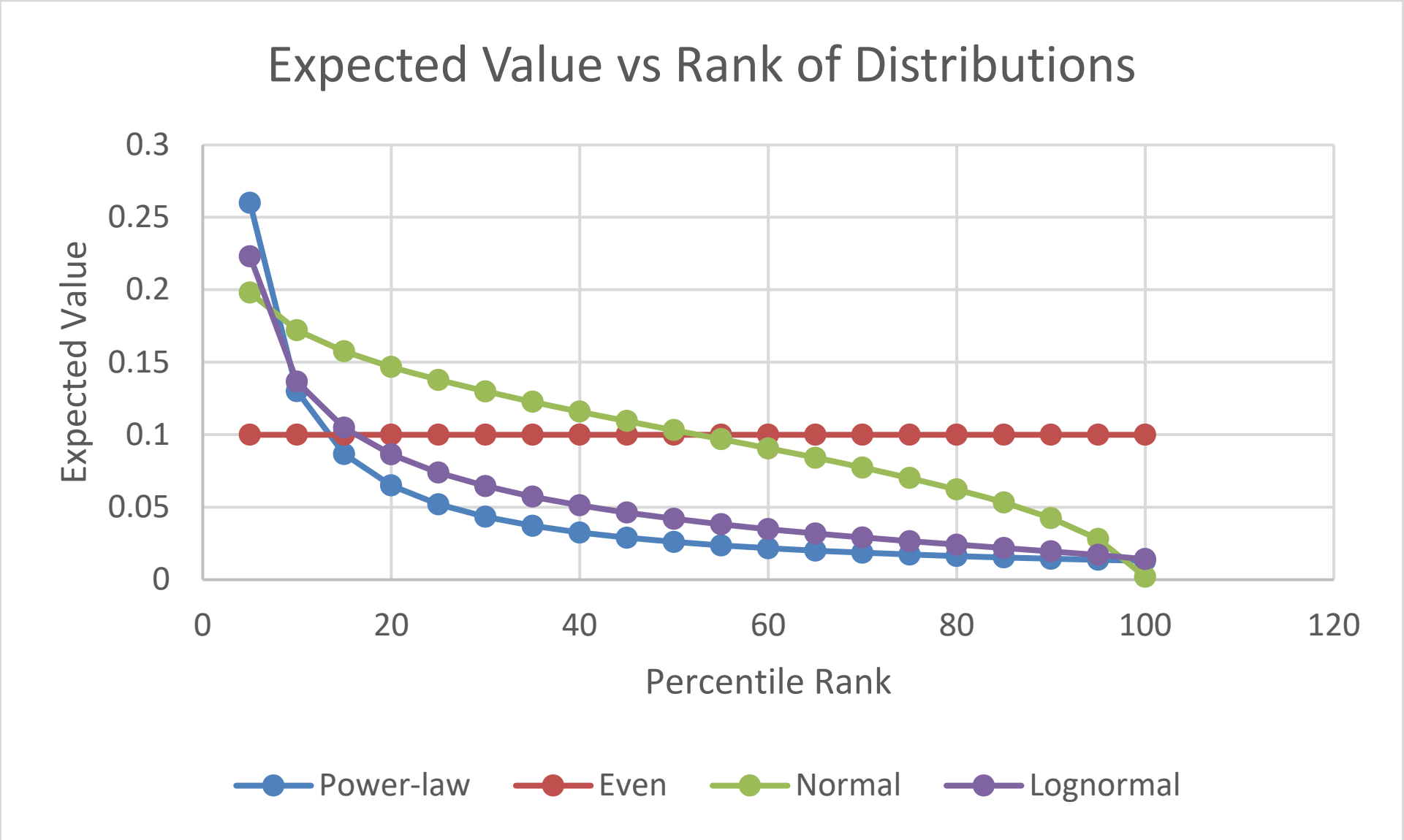

Figure 1b shows that in expected value vs rank, these distributions can look similar, unlike in PDF plot of Figure 1a. Normal distribution is more confined in a range of expected values than lognormal and power-law distributions, except at extreme tails. Power-law has a fast decaying expected value as rank increases. Lognormal distribution sits somewhere in between and looks similar to power-law. It resembles power-law at the extreme of large values while it is also bounded at a minimum value.

**General Concentrated Distributions**

While power-law distributions provide a critical framework for analyzing hierarchical systems, general concentrated distributions offer a more comprehensive perspective, particularly for dynamic systems undergoing continuous change. Beyond power-law, other concentrated distributions such as exponential and Poisson distributions share common traits of convexity and rapidly declining probabilities. These transient states often emerge in dynamic systems and offer broader applicability in understanding real-world phenomena.

There are other concentrated distributions, which do not follow power-law exactly. These are transient states and are not scale-free. These general concentrated distributions also have a PDF *f(x)* that drops as *x* increases ($\frac{\partial f(x)}{\partial x} < 0$ ) and have a convex shape ($\frac{\partial^2 f(x)}{\partial x^2} > 0$). Hence, *f(x)* peaks at minimum value of *x* and drops faster at lower *x* values than at high *x* values. For example, exponential distribution $f(x) \sim e^{-\lambda x}$ for continuous *x* variables and Poisson distributions $f(x) \sim \lambda^x / x!$ for large discrete *x* variables. These properties are also true in rank space when looking at the curve of expected value vs rank.

Lognormal distributions transition asymptotically into power-law at large values, underscoring their relevance in modeling dynamic growth systems. In fact, this is the core pillar of random growth model explaining how power-law emerges in a linear growth model.

General concentrated distributions are more common than power-law. In most of the cases, the variables are going through dynamic changes. Hence, studying general concentrated distributions have more applications than power-law.

**Real Life Example: Income and Wealth Distributions**

Let us look at some real world data. Income and wealth distributions in the United States and Japan provide illustrative examples of power-law and lognormal behaviors. U.S. income and wealth distributions exhibit steeper power-law characteristics compared to Japan, reflecting differences in socio-economic structures, education, and labor practices. As indicated by the Gini Index with the U.S. at 0.45 and Japan at 0.33, variations in systemic factors influence these distributions.

Figure 2. US and Japan Income Distribution in Percentile Ranks

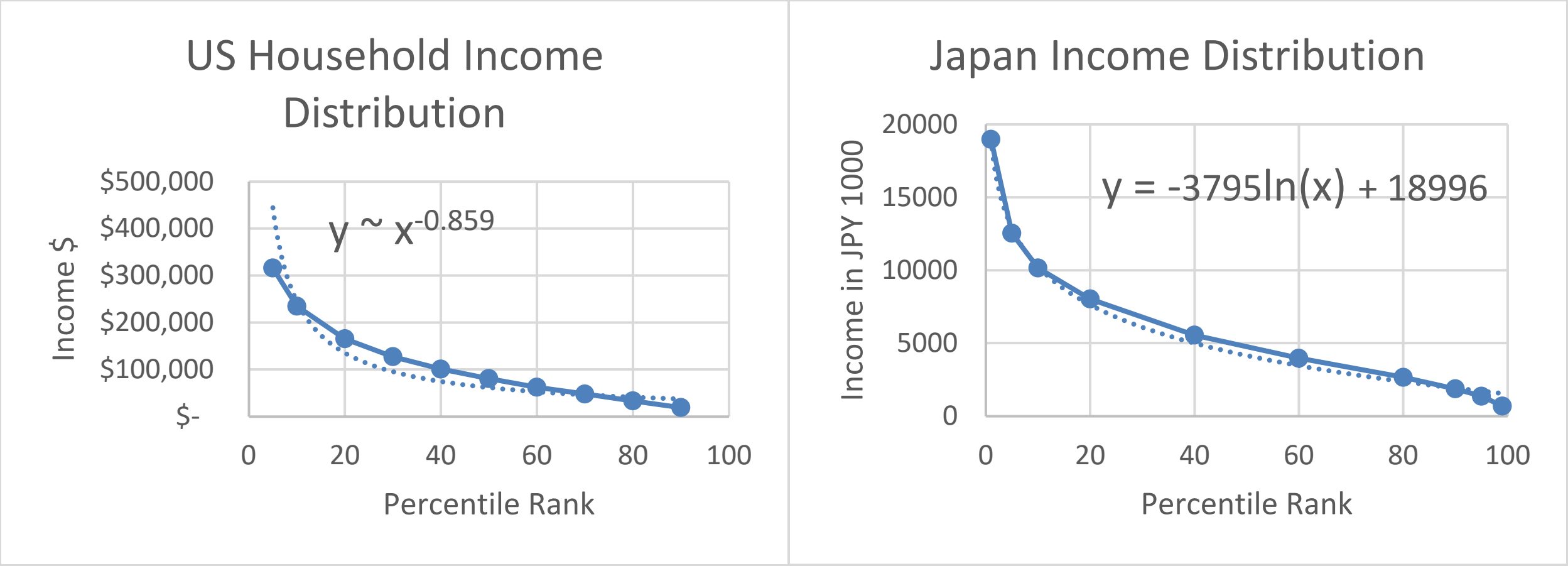

Figure 2 shows 2023 US household income distribution using government census data (census.gov, Table A-4a). Income distribution in the US follows power-law approximately with a slope in log-log plot of -0.86. The 2014 Japan income distribution data is from a research paper [Kitao 2019, Table 3]. Japan income distribution fits closer to a lognormal distribution.

As we discussed above, the PDF plots can reveal more information of the distributions.

Figure 3. Comparison of US and Japan Income Distributions in PDF

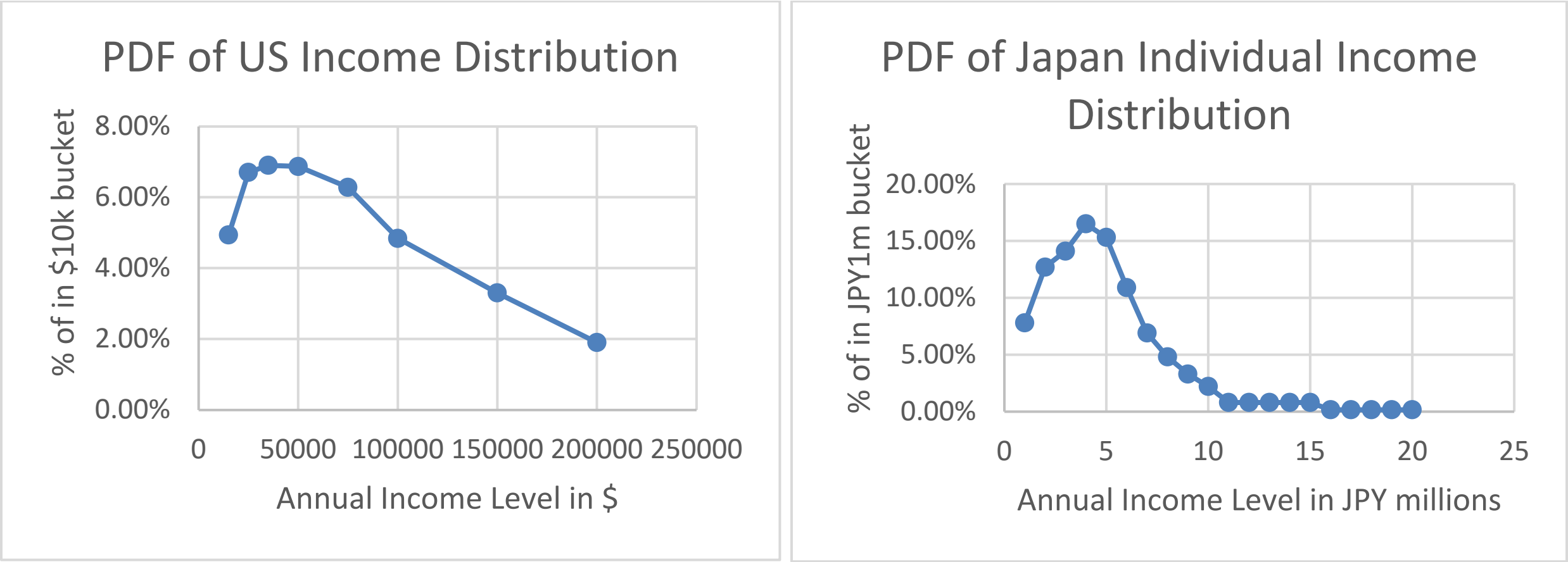

Figure 3 shows the PDF plots of income distributions in the US and Japan. The US data is published by US Census Bureau of US Dept. of Commerce for 2023 US household income. The original income buckets were not evenly spaced. To derive PDF, we calculated the CDF at each income level and then normalized to equal income buckets by taking the derivatives of CDF. The vertical axis is the percentage of US households in $10,000 bucket. The horizontal axis is the annual income level. The data for Japan is from Japan National Tax Agency (nta.go.jp)'s 2022 individual income Table 16. The original data is already bucketed of JPY 1 million. Vertical axis is the percentage of individuals in each bucket. Horizontal axis is the annual income levels.

Presented in PDF, US income distribution is not exactly power-law. Its peak probability density occurs in middle class in the US income distribution. In PDF, Japan's income distribution consistently follow a lognormal distribution as in percentile rank. Its middle class size is bigger than the US reflecting more economically equal society [Arizawa 2017]. Income differences in Japan are mainly from factors such as education and jobs at large companies vs irregular workers, while factors of education, racial differences in wages, and working hours explain most of the differences in the US [Guzman 2024].

Next, let us look at the wealth distributions, which is expected to be more concentrated than income distributions, because of the cumulative effect of wealth over time. Wealth data is in general less available than the income data.

Figure 4. US and Japan Wealth Distributions in Percentile Ranks

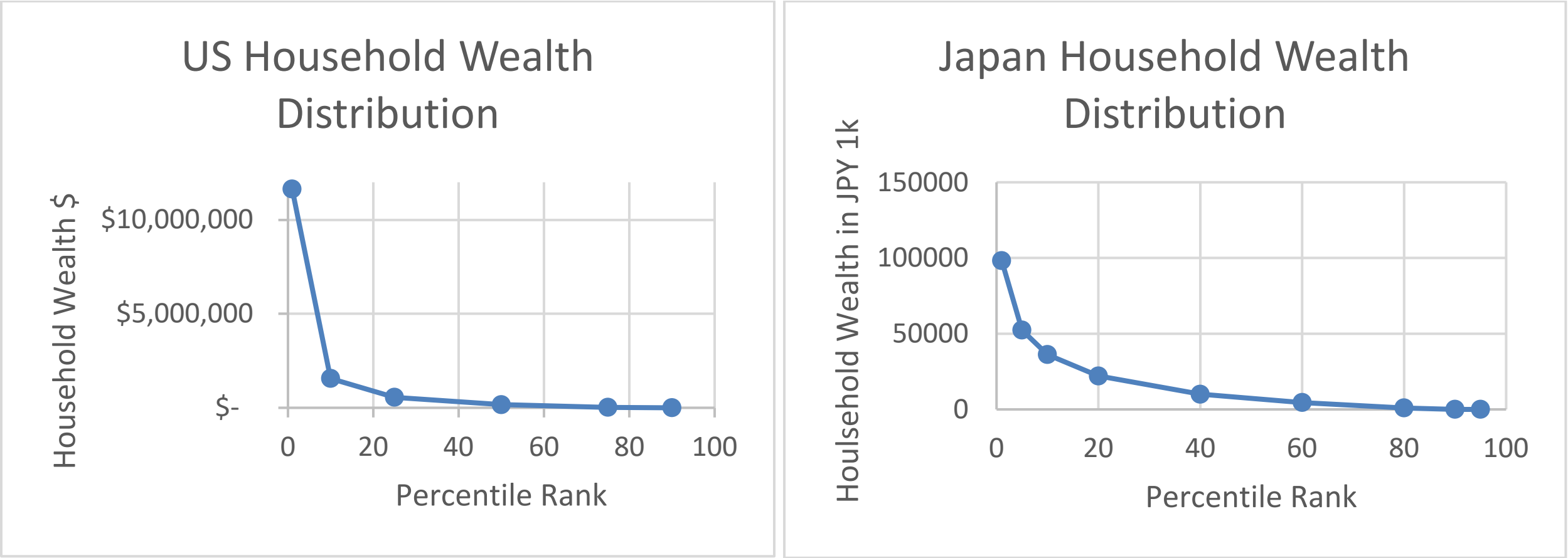

Figure 4 shows the US household wealth distribution from 2022 Richmond Fed survey. The Japan household wealth distribution data is for 2014, presented in a research paper [Kitao 2019, Table 4]. Both the US and Japan distributions fit power-law, with log-log slope parameter for Japan about 2.0, not as steep as the US's 2.4. This is consistent with the general understanding that Japan's wealth distribution is less concentrated than in the US [Saez 2014].

Figure 5. Comparison of US and Japan Wealth Distributions in PDF

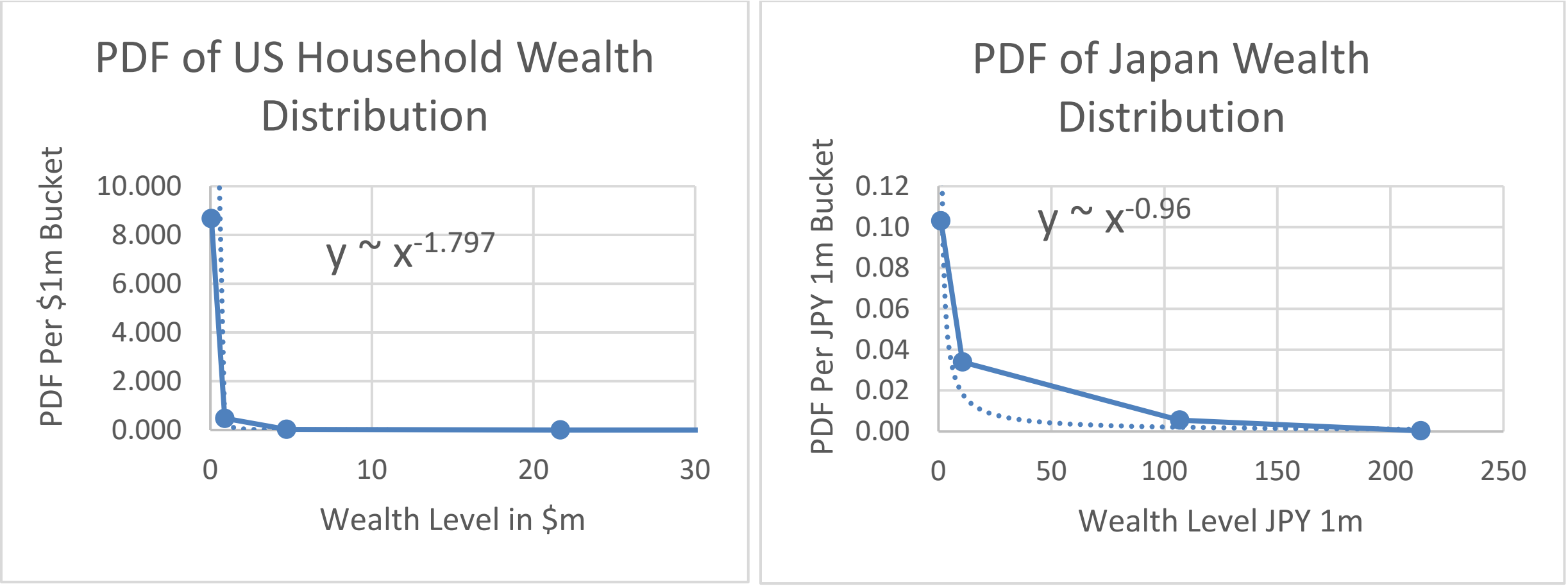

Figure 5 compares the PDF of wealth distribution of the US and Japan. The US data is from Federal Reserve Economic Data by Federal Reserve Bank of St. Louis for 2024 Q2. Data for Japan is from online (statista.com) for 2020. Survey data often has percentages of population associated with non-uniform wealth buckets. The PDF data is calculated by taking derivative of CDF data and normalized for equal wealth buckets.

As we can see, wealth distribution in both countries follow power-law, not lognormal [Kitao 2019]. The log-log slope in the US is significantly steeper than Japan, reflecting higher inequality in the US than Japan.

The obvious question is what caused the differences between Japan and the US income and wealth distributions. While this paper is not about macroeconomics, in the next sections of this paper, I will investigate purely from how multi-agent large systems work via a feedback loop model. It turns out; the key differentiator separating agents is their response functions to observations.

## III. Modeling Self-Organized Crowd with Feedback Loop

In this section, I will recap the model where interaction of agents, observation, and external force is modeled as a feedback loop. At each time step, agent's incremental change of decision $dS_i$ is determined by its responses to external force change $dE$ and observation change $dO$, plus some random noise $\varepsilon_i$.

$$dS_i = C_i\, dE + B_i\, dO + \varepsilon_i \tag{1}$$

where $C_i$ is the response coefficient to dE, $B_i$ is the response coefficient to dO.

In equation (1), $\varepsilon_i$ is a stochastic term including random noise of Wiener process and drift, describing an agent's decision not determined by $dE$ or $dO$. For simplicity, in later discussion, we often assume the impact of $\varepsilon_i$ is small compared with the first two terms and drop $\varepsilon_i$ in modified equation (2). The term $\varepsilon_i$ is important when correlation between agents is considered [Xia 2016], e.g. for calculating order parameter for synchronization. The drift term $\mu\, dt$ represents the agent's biases.

$$dS_i = C_i\, dE + B_i\, dO \tag{2}$$

### Response Function

Agents respond to the change of observation via different coefficients $B_i$, which are the response functions of agents. The model does not require only one central observation. There can be local observations by a few neighboring agents and local feedback loops. Agents can have impact only on local observation, i.e. $B_i = 0$ except for local observation. The observation can also be global by all agents.

In financial markets, $dS$ is people's decision to buy or sell certain securities. $dO$ is the price change. $dE$ is the news outside of price change. When $B_i>0$, the player is a momentum follower; when $B_i<0$, the player is a contrarian.

Agents' power to influence the outcome is denoted as $A_i$. This paper extends from previous model [Xia 2016] and makes $A_i$ vary across agents and time steps. Distribution of $A_i(t)$ is the focus of this study.

Change of observation is determined by the aggregated decisions by agents weighted by their power or wealth $A_i$. N is total number of agents.

$$dO = \sum_{i=1}^{N} A_i dS_i \tag{3}$$

We can also normalize agents' wealth and present as percentage share of total power,

$$\sum_{i=1}^{N} A_i = 100\% \tag{4}$$

Let us now introduce time dependence, from time *t-dt* to *t*, Equation (3) becomes,

$$dO(t) = \sum_{i=1}^{N} A_i(t)\, dS_i\,(t) \tag{5}$$

Agent's action $A_i(t)\; dS_i(t)$ at time t has an instantaneous impact on the outcome *dO(t)*. In Equation (2), agent's decision at *t* is influenced by previous observation at *t – dt* and current external force.

$$dS_i(t) = C_i\,(t)\, dE(t) + B_i\,(t)\, dO(t - dt) \tag{6}$$

Hence,

$$dO(t) = \sum_{i=1}^{N} A_i(t)[C_i\,(t)\, dE(t) + B_i\,(t)\, dO(t - dt)\,] \tag{7}$$

It is assumed $C_i$>0 and agents respond positively to external force *dE*. If the external force persists in the same direction, most agents follow momentum and observation trends in the same direction.

If all agents have $B_i$=0 and $A_iB_i$=0, there is no endogenous reaction, observation simply responds to exogenous signals *dE*.

When there is no external force, *dE(t)*=0, and $B_i$=0, all agents have no action, $dS_i$=0, agents have no impact on observation *dO*. In this case, random noise term $\varepsilon_i$ in equation (1) plays more important role in determining *dO*.

If $B_i$ is not zero,

$$dO(t) = \sum_{i=1}^{N} A_i(t) B_i\,(t)\, dO(t - dt) \tag{8}$$

Equation (8) describes an isolated many-body system. The crowd behavior is simply determined by its endogenous properties, or the feedback parameters $A_i$ and $B_i$. We define

$$D(t) = \sum_{\boldsymbol{i=1}}^{\boldsymbol{N}} A_i(t)B_i\,(t) \tag{9}$$

$$dO(t) = D(t) * dO(t-dt) \tag{10}$$

If *D* is a constant at all time steps,

$$dO(ndt) = dO(0)\, D^n \tag{11}$$

*D*>0 means that the weighted response of the whole many-body system is to follow the direction of the previous observation *dO(t-dt). D*<0 means that the weighted response of the system is to reverse the direction of the previous observation. *D*>1 means $dO(t)$> $dO(t-dt)$ , that the weighted response is not only to follow the direction of the previous observation, but also amplify it. The system can compound on itself when *D*>1 and *dO* will become unbounded.

Agents with disproportionally large wealth are called super agents. For example, if agent 1 is the only super-agent, $A_1$>>$A_i$ for all other *i*'s, then *D* is approximately $A_1B_1$.

## IV. Reward Function of Agents' Wealth

Response Function was shown as a key driver of synchronization [Xia 2016]. In this section, I focus on the reward function for change of wealth of agents.

The process of wealth $A_i$ change over time depends on specific reward mechanisms. For example, tossing a coin can have independent wages at each time step. Holding a stock position can have profit/loss (p/l) from cumulative position sizes with p/l reinvested in the stock at each time step. Network's nodes add new links based on how many links the node has acquired till that time vs the total number of links [Barabasi 1999]. To specify how $A_i(t)$ change over time, we need to define reward function for specific problems.

When agents acted correctly, $dS_i\,(t)$ and $dO(t)$ have the same signs, $A_i$ increases. Otherwise, $A_i$ decreases. Wealth at $t$+$dt$, $A_i(t+dt)$ = $A_i(t) + dA_i(t)$, where $dA_i(t)$ is the reward obtained at time *t*. Reward $dA_i(t)$ is determined by agent's action $A_i(t)\, dS_i\,(t)$ and outcome *dO(t),* i.e. $dA_i(t)$ is a function of $A_i\,(t), dS_i\,(t), dO(t)$.

From equation (6), when *dE*=0, $dS_i(t) = B_i\,(t)\, dO(t-dt)$. When all agents start with the same wealth at time 0, response function $B_i$ is the differentiator that leads to distribution of $A_i(t)$ across agents over time.

Let us look at a special case.

## Binary Outcomes in Observations

Tossing a coin has only two binary outcomes, head or tail. Only the sign of $dO(t)$ matters. $Sign[dO(t)]$ = +1 or -1. $dS_i$ is the incremental change of decision in either direction. $dS_i(t) = B_i\ (t)\ sign[dO(t-dt)]$. It can be positive or negative. For convenience, we confine $-1 \leq B_i \leq 1,\ -1 \leq dS_i \leq 1$ . Wealth becomes

$$A_i(t+dt) = A_i\ (t) + A_i\ (t)\ dS_i\ (t)\ Sign[dO(t)] \tag{12}$$

Or reward function is

$$dA_i(t+dt) = A_i\ (t)\ dS_i\ (t)\ Sign[dO(t)] \tag{13}$$

Note the change of wealth, $dA_i$ is a linear function of $A_i$. This is an important condition for power-law to occur, as formulated in random growth model and network preferential attachment model. Total wealth (or number of links in a network) grows linearly with time *t*.

Random growth model has the growth rate randomly drawn, leading to lognormal distribution. Only for very large $A_i$ values, it approaches power-law. Boundary conditions of $A_{i_min}$ is also critical for power-law to occur.

In order for power-law to occur, the reward function needs to be close to linear but with some deviation or boundary conditions such as imposing a minimum value. This indicates that power-law is a special case of many possible concentrated distributions. Properties of power-law, such as scale-free, are rather unique.

Since *dO(t)* is dependent on $A_i(t)$, $dA_i$ is not exactly linear to $A_i$. Only in special cases, Ai will follow power-law distribution. Otherwise, the distribution is not stationary as it keeps evolving.

From equation (2) when *dE*=0 and $B_i$ is time-invariant,

$$dS_i(t) = B_i\ Sign[dO(t-dt)] \tag{14}$$

From equation (5),

$$dO(t) = Sign[dO(t-dt]\ \ \sum_{\boldsymbol{i=1}}^{\boldsymbol{N}} A_i(\mathrm{t})B_i \tag{15}$$

Hence

$$Sign[dO(t)] = Sign[dO(t-dt]\ Sign[\sum_{\boldsymbol{i=1}}^{\boldsymbol{N}} A_i(\mathrm{t})B_i\ ] \tag{16}$$

Now equation (12) can be rewritten by using equations (14-16),

$$A_i(t+dt) = A_i\,(t)\{1 + B_i\,Sign[dO(t-dt)]\,Sign[dO(t)]\}$$
$$= A_i\,(t)\{1 + B_i\,Sign[\textstyle\sum_{i=1}^{N} A_i(\mathrm{t})B_i\,]\} \qquad (17)$$

In a trendy move, trend-following agents with $B_i > 0$ can accumulate larger wealth. Hence, $Sign[\sum_{i=1}^{N} A_i(\mathrm{t})B_i\,] = 1$.

$$A_i(t+dt) = A_i\,(t)(1+B_i) \qquad (18)$$

$$A_1(ndt) = A_1\,(0)\,(1+B_1)^n \qquad (19)$$

$$A_i(ndt) = A_i\,(0)\,(1+B_i)^n \qquad (20)$$

If all agents start with the same wealth at time 0, after normalization making $\sum_{i=1}^{N} A_i(\mathrm{t}) = 1$,

$$A_i(ndt) = (1+B_i)^n \,/\sum_{i=1}^{N}(1+B_i)^n \qquad (21)$$

$$A_i(ndt) = A_1(ndt)\,(\frac{1+B_i}{1+B_1})^n = A_1(ndt)\,\alpha_i{}^n \qquad (22)$$

Where,

$\alpha_i = \frac{1+B_i}{1+B_1}$ . Let us assume -1=<$B_i$ =<1, and $B_i$>$B_{i+1}$ for any i, or $B_i$ is a declining function of i, $dB_i/d\,i$ <0. Hence, $\alpha_1 = 1$, $0 < \alpha_{i+1} < \alpha_i < 1$, for all $i > 1$.

Next, let us discuss several possible distributions emerging from this generic model.

**General Concentrated Distribution**

From equation (22), we can easily show that $\frac{\partial A_i}{\partial i} < 0$ and $\frac{\partial^2 A_i}{\partial i^2} > 0$, if time step $n$ is big enough. Hence $A_i$ decreases as i increases in a convex shape of curve. Such function $A_i$ is not always exactly power-law. It can be exponential functions or other forms. $A_1$, as the highest ranked agent (the wealthiest), has much higher share of wealth compared with poor agents.

**Effect of Boundary Conditions without Feedback Loop**

When there is no feedback loop and all agents have $B_i$=0, $dS_i(t) = 0$. Observations $dO(t)$ changes randomly. Let us assume the reward function $dA_i(t+dt) = \varphi$ , which follows i.i.d. (independently and identically distributed) random walks. Starting with even distribution, i.e. all agents have the same wealth at time 0, $A_i(t)$ evolves to normal distribution center around original wealth level.

If we impose boundary conditions such as a minimum wealth level, e.g. $A_i(t) \geq A_{min}$, the distribution of $A_i(t)$ eventually becomes concentrated with peak probability (most of the agents) at the boundary with minimum wealth. The PDFs of distribution of wealth of agents at different time steps are numerically simulated. Similar results were obtained with different assumptions in reward functions [Dragulescu 2000, physics.umd.edu website].

Next, let us assume the reward function is linear to $A_i$, $dA_i(t+dt) = A_i(t)\varphi$ where $\varphi$ follows i.i.d. random walks. This is the setup of Random Growth model. Without boundaries, distribution of $A_i(t)$ becomes lognormal over time. When boundary condition of $A_i(t) \geq A_{min}$, is imposed, the distribution of $A_i(t)$ approaches power-law for large values of $A_i$.

**Case of Power-law Distribution**

In a special case, $A_i(t)$ follows exactly power-law distribution. When $1 + B_i = c/i^b$, where c and $b$ are constants. $B_i = \frac{c}{i^b} - 1$ is a declining function of *i*.

From equations (20) and (22),

$$A_i(ndt) = A_i\ (0)(1+B_i)^n = A_i\ (0)c^n/i^{bn} \tag{23}$$

$$\frac{A_i(ndt)}{A_1(ndt)} = \left[\frac{A_i(0)}{A_1(0)}\right]/i^{bn} \tag{24}$$

$$A_i(t) \sim 1/i^{bn} \tag{25}$$

It follows Zipf's law in expected value vs rank, which satisfies power-law distribution for all values of wealth.

**Linear Declining $B_i$**

In another special case, N agents start with even wealth distribution, $A_i$(t=0)=$A$ constant for all agents. $B_i = 1 - 2\frac{i-1}{N-1}$, is a linear declining function of *i*. $B_1$=1 and $B_N$= -1. At mid-point *i=(N+1)/2*, *(i-1)/(N-1)=1/2*, $B_i = 0$.

Equation (20) becomes

$$A_i(ndt) = A_i\ (0)(1+B_i)^n = A_i\ (0)2^n\left(\frac{N-i}{N-1}\right)^n$$

Hence, $A_1(ndt) = 2^nA$. $A_N(ndt) = 0$. At mid-point *(i-1)/(N-1)=1/2*, $A_i(ndt) = A$ =$A_i\ (0)$. No change from *t=0*.

Here $A_i(t)$ does not follow power-law distribution. Instead, it follows a general concentrated distribution where Matthew Effect is also apparent. As the crowd system continues to evolve, further concentration happens. Distribution has a faster decay of expected value vs rank, or slow decay in probability density function for large values, i.e. extreme fat tail. Most agents are poor with minimum wealth, while a few agents have very large wealth. This is not a steady state. Super agents will become monopoly. Eventually only external forces can break up the dominance.

Figure 6. Simulated Wealth Distribution

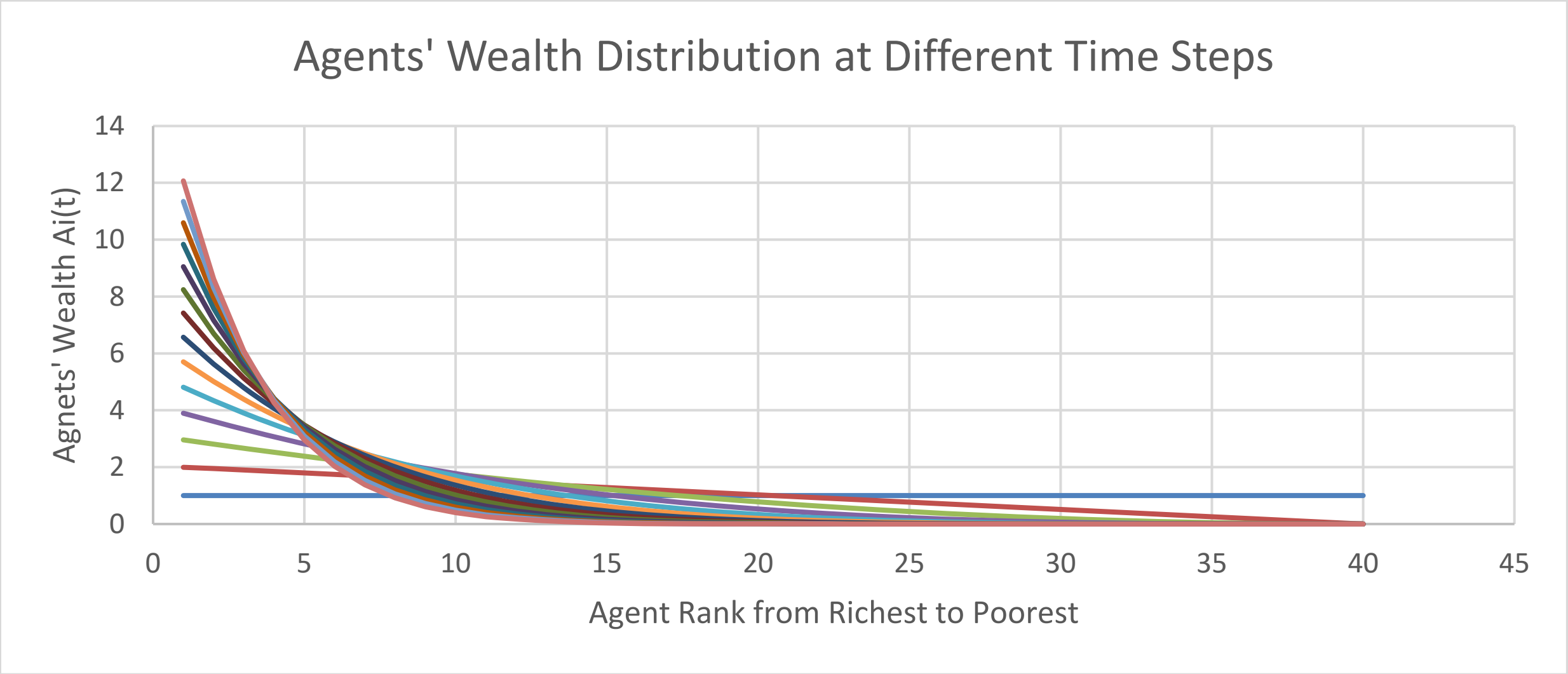

Figure 6 shows how agents' wealth change over time. Starting with even wealth distribution for all agents at time 0, wealth shifts to concentrate with agents who have high response functions.

## V. Discussions

To summarize, there are several causes of power-law and general concentrated distributions, 1) there is a feedback loop; 2) there is an external force to start a trend; 3) agents' different response functions separate them in the final wealth distributions.

In the case of dynamic adapting agents, $B_i$ can change over time. Like discussed in my previous paper [Xia 2016], agents tend to move from normal state to a more reactive state when there is a strong trend. The agents with high $B_i$ values benefit more from these trendy moves. When all other agents adapt to the same value as the highest $B_i$ value, they do not gain more relative wealth share. They simply maintain their wealth share. If they stay with low $B_i$ values, the relative wealth will continue to decline. In a trendy market, if an agent gets the direction right but acts with small size, he still loses his share of total wealth. This is why FOMO (fear of missing out) is equivalent to fear-of-loss when lagging agents lose relative share of wealth even though their nominal wealth increases. Their relative purchasing power decreases.

In the example of wealth distribution, rich people gain more absolute dollar value when they compound at the same return as other people. More importantly, rich people have more disposal income and higher risk tolerance for investments. They have higher response function $B_i$ to the observation. When market has big trending moves, they benefit more by investing higher percentage of their wealth in the market.

There are real-world examples where distributions transition from normal or lognormal to a power-law as systems evolve or experience changes. Such transitions often occur due to underlying shifts in dynamics of $B_i$, such as network effects, preferential attachment, or systemic feedback loops. Here are some examples:

1. Income and Wealth Distribution

In early stages of industrialization or economic development, wealth or income distribution often approximates a lognormal distribution, as the feedback loop is weak. For example, China in 1970s (precise data is not available) had relatively low inequality, while the overall wealth was very low. As economic systems mature and inequality rises, wealth distribution shifts to a power-law distribution, where a small number of individuals hold the majority of wealth. Over time, wealth accumulation by the top earners led to power-law behavior, as observed in modern-day Chinese wealth distribution. Better linked (higher response functions) society produces higher total wealth, but also higher inequality. Is this an inevitable outcome of economic and technological advancements? I leave this important question for economists to answer.

In our earlier comparison of wealth distributions in the US and Japan, one possible explanation for the differences is that people in Japan are likely to have more confined and lower values of their response functions of $B_i$.

Figure 7. US Household Wealth Distribution from 1989 to 2024

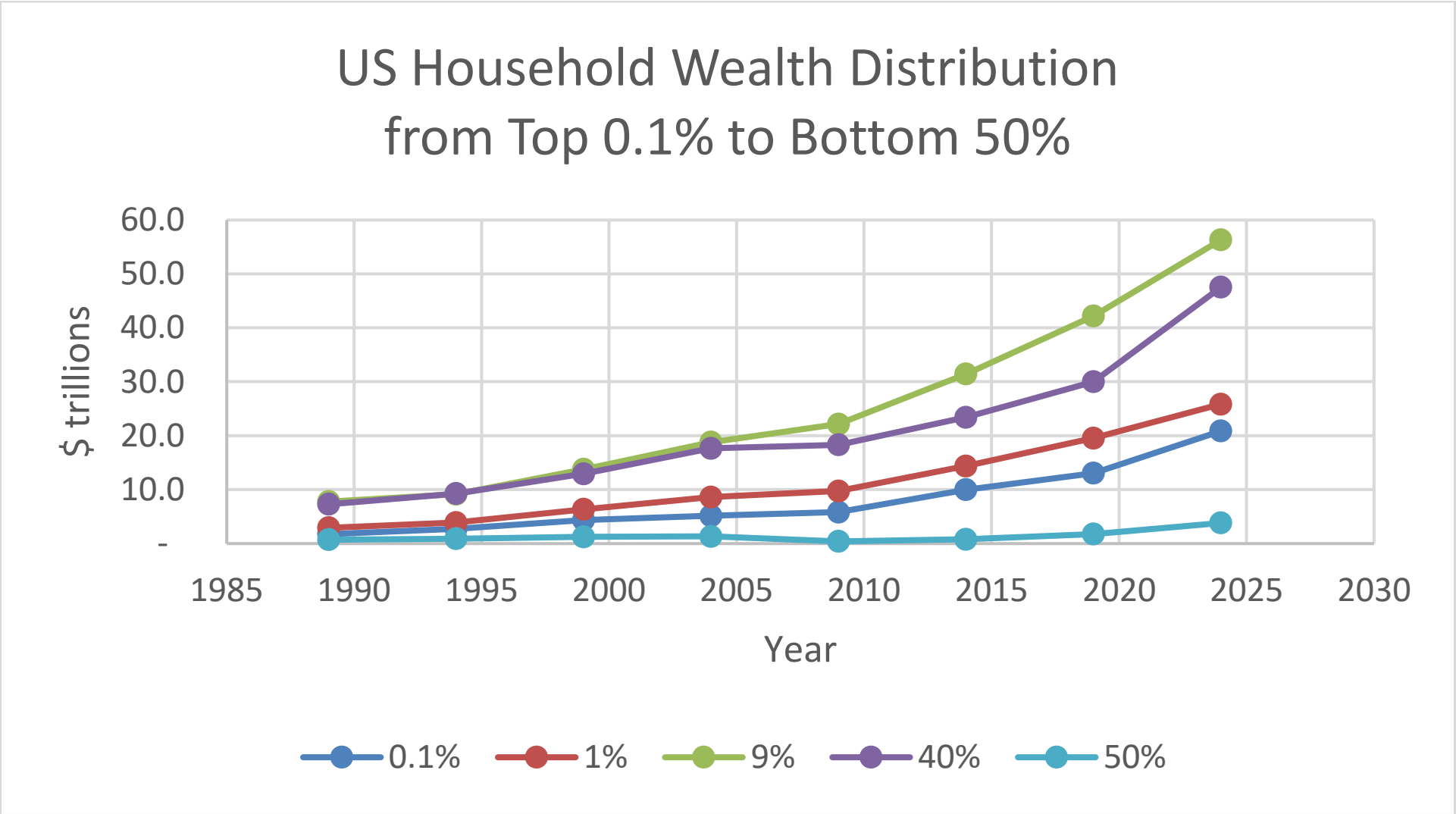

Figure 7 shows the change of US household wealth distribution in the last 35 years. The data is from Federal Reserve Economic Data by Federal Reserve Bank of St. Louis for 2024 Q2 update. Total wealth is bucketed in top 0.1%, 1%, 9%, 40% and bottom 50% of the households.

Total wealth of bottom 50% of the US households lost 85% during the 2008-2010 global financial crisis (GFC). It took 12 years till 2017 to recover to the pre-GFC high watermark in 2005. All higher buckets had a relatively steady increase of their wealth from 1989 to 2024. Total overall wealth increased from $20.9 trillion in 1989 to $154 trillion in 2024. The richest households in the population gained more relative wealth share in the past 35 years.

The power-law phenomenon happens more during crisis or a strong trend. VC funds is a good example where gains are big when trend is strong [Mallaby 2022].

2. <u>City Sizes</u>

In the early stages of urban development, city sizes may follow a normal or lognormal distribution due to random growth processes. As cities grow and urban migration increases, city size distribution shifts toward a power-law, governed by Zipf's Law, where a few megacities dominate the population. For example, during early urbanization phases in countries like India or China, city sizes were lognormally distributed. As urbanization advanced, the distribution shifted, with cities like Mumbai and Shanghai emerging as outliers following a power-law distribution. This can be explained by higher response functions, which reflect stronger linkage of cities and migration flows.

3. <u>Market Returns and Financial Crises</u>

In stable market conditions, asset returns are often modeled as normal or lognormal distributions. During periods of market stress or crises, extreme events dominate, leading to fat tails and power-law behavior in return distributions. For example, the 2008 financial crisis showed a shift in asset return distributions, where extreme drawdowns occurred more frequently than predicted by normal models.

By monitoring the response functions $B_i$ of market participants, we can forecast financial crisis. This is now becoming more possible as more data is available in social media and brokerage accounts. Applications of AI to predicating financial crisis and bank runs are not in a distant future [Shiller 2015, Lo 2024]. This model provides a logical explanation of where to look for the critical frontier data.

4. <u>Company Sizes and Market Capitalizations</u>

In new industries or markets, company sizes may initially follow a lognormal distribution due to organic growth. As industries mature and network effects (e.g., preferential attachment) take hold, company sizes follow a power-law distribution. For example, in the early days of the tech industry, company valuations were relatively uniform. Today, companies like the “Magnificent Seven” dominate the market, and firm sizes exhibit power-law characteristics

5. Scientific Citations

In emerging research fields, citation counts for academic papers often follow a lognormal distribution. Over time, certain papers gain preferential attachment (e.g., being cited because they are already widely cited), resulting in a power-law distribution. For example, the early distribution of citations for papers in computer science was lognormal. Over time, landmark papers like Google's PageRank transitioned the citation network to a power-law.

6. Online Content Popularity

In the early stages of a platform's life cycle, content views or shares are often lognormally distributed. As platforms grow, network effects, recommendation algorithms, and viral dynamics lead to a power-law distribution, with a few pieces of content dominating. For example, in early YouTube, video views were lognormal. Today, videos like "Despacito" have billions of views, following a power-law distribution.

7. Collective Intelligence

This model can be used to study other crowd behaviors. For example, collective intelligence requires agents act independently [Malone 2022], where minimum synchronization among the agents is essential to ensure independent thinking. Synchronization and concentration can turn crowd wisdom to crowd madness. Majority's views can overtake the crowd, as minority feel insecure to express their views. For example, in job interviews, candidates tend to conform their views to the interviewer's in order to get the offers. The organization ends up hiring people with similar views.

8. Inverse Problem of Design and Systems of AI Agents

If there is a desired distribution of wealth, we can reverse engineer and design the system with optimal response functions of $B_i$ from government super agents such as the Fed and other central banks. In the future, if more tasks are completed by artificial intelligence agents, we can potentially program the response functions into the agents to achieve system level objectives. Macroscopic parameters such as entropy, energy storage, stability, and order offer insights into system-level dynamics. These topics, while beyond the scope of this paper, highlight the potential for further exploration into multi-agent systems and their macroscopic properties [Xia 2013].

## VI. Conclusions

This study introduces several novel insights into the understanding of power-law distributions of wealth in a crowd.

1. Emergence of power-law and other concentrated wealth distributions is explained by a feedback loop model for multi-agent crowd.

2. A unified representation of probability density function (PDF), cumulative distribution function (CDF), and rank is essential for better comprehension of power-law phenomena.
3. While power-law distributions are frequently associated with the upper tail comprising extremely large values, the lower tail of small values provides a more critical perspective. The peak in the PDF serves as an effective indicator to distinguish between normal or lognormal distributions and power-law or general concentrated distributions.
4. Power-law, as a rare steady state, generally manifests in the asymptotic behavior of large values but can occasionally extend across the entire range of values.
5. The emergence of power-law in random growth and network models necessitates a linear reward function; however, perfect linearity alone is insufficient. Specific boundary conditions are essential for its realization.
6. The response function $B_i$ is a distinguishing factor among agents, playing a crucial role in determining the distribution of wealth. Collecting data on $B_i$ is critical for monitoring multi-agent systems, such as forecasting financial crises or predicting viral trends.
7. Differences in response functions may explain the variations in income and wealth distributions between the United States and Japan. Total societal wealth is driven by healthy competition, which, while fostering growth, inherently leads to income and wealth inequality. Inequality, as a byproduct of competition, is not inherently negative provided there is sufficient social mobility. Governments, acting as super-agents, may influence the relative distribution of wealth while maintaining incentives for overall growth.

ACKNOWLEDGEMENTS: The author would like to thank the following people for helpful discussions [To be completed].